\documentclass[11pt,twoside]{article}
\usepackage{asp2010}

\resetcounters

\begin{document}

\title{The Astrophysics Source Code Library: Where do we go from here?}
\author{Alice Allen$^1$, Bruce Berriman$^{2,3}$, Kimberly DuPrie$^1$, Robert J. Hanisch$^{4,3}$, Jessica Mink$^5$, Robert Nemiroff$^6$, Lior Shamir$^7$, Keith Shortridge$^8$, Mark Taylor$^9$,  Peter Teuben$^{10}$, and John Wallin$^{11}$
\affil{$^1$Astrophysics Source Code Library}
\affil{$^2$Infrared Processing and Analysis Center, California Institute of Technology}
\affil{$^3$Virtual Astronomical Observatory}
\affil{$^4$Space Telescope Science Institute}
\affil{$^5$Harvard-Smithsonian Center for Astrophysics}
\affil{$^6$Michigan Technological University}
\affil{$^7$Lawrence Technological University}
\affil{$^8$Australian Astronomical Observatory}
\affil{$^9$University of Bristol}
\affil{$^{10}$University of Maryland}
\affil{$^{11}$Middle Tennessee State University}}

\begin{abstract}
The Astrophysics Source Code Library\footnote{\url{http://ascl.net}}, started in 1999, has in the past three years grown from a repository for 40 codes to a registry of over 700 codes that are now indexed by ADS. What comes next? We examine the future of the ASCL, the challenges facing it, the rationale behind its practices, and the need to balance what we might do with what we have the resources to accomplish.

\end{abstract}

\section{Introduction}
The Astrophysics Source Code Library, or ASCL, is a free online registry for source codes used in astronomy and astrophysics. It exists to make codes more discoverable for examination and potential re-use. We also hope to foster more recognition of the contributions codes and their authors make to the community. The guiding principle is increasing the integrity and reproducibility of research by increasing the transparency of computational research methods.

\section{History and Activities of the ASCL}
The ASCL was started in 1999 by Robert Nemiroff and John Wallin \citep{RJN_aas1999}. In 2010, the resource was moved to the PHP Bulletin Board (phpBB) hosting the Astronomy Picture of the Day (APOD) discussion forum, and became primarily a registry rather than a repository. Nemiroff enlisted Alice Allen's help; after moving the existing entries to the new site, she started adding new ones in the role of ASCL editor. The ASCL retains the ability to house programs, but does not require they be deposited. 

In early 2011, Peter Teuben and Allen formed the Advisory Committee for the ASCL; members are listed as coauthors of this article. Later that same year, Kimberly DuPrie came on board as associate editor. 

In January 2012, ADS started indexing ASCL contents. This makes codes more discoverable, and because they have stand-alone identifiers, ADS listings for papers using codes can have links to the software entries. Indexing by the ASCL and ADS also provides a way for codes that do not have papers describing them to be cited and have those citations tracked. Over the past three years, the ASCL has grown from about 40 codes to over 700. When looking at hits on the ASCL for a three-week period in March 2013\footnote{\url{http://asterisk.apod.com/wp/?p=309}}, we found that 60\% came from outside the US and that in that three-week period, 83 countries had accessed the ASCL. 

Our goal is this: when you read a paper and want to see the code used, you click a link or two and can look at the program for its underlying assumptions, methods, and computations. If you want to investigate an unfamiliar domain, you can search the ASCL to see what codes have been written in that area. To this end, we are active in fostering code sharing; we have organized Birds of a Feather sessions \citep{B02_adassxxiii}, and with the American Astronomical Society's Working Group on Astronomical Software, co-sponsored a session at AAS meetings on the topic. The ASCL is starting to have an impact in the community \citep{P069_adassxxiii}.

\section{A Typical Entry}
An entry consists of five fields:
\begin{itemize}
\item code name
\item description of the software; this was originally a copy of the abstract for the code paper, but more recent entries are brief descriptions of the software instead.
\item name of the author(s)
\item URL for the program's website or download location
\item unique identifier assigned by the ASCL, the ASCL ID. This number reveals the month and year the code was assigned an ID and is used to build the permalink for the code, just as astro-ph does for preprints.
\end{itemize}

We also usually include a link to a paper describing a code or using the program for additional information.

\section{Practices and Rationale}
We register software rather than serve as a repository because authors like to keep their codes close to them; we can act as a repository, however, and welcome code deposits. We keep the metadata light because we point to other places where the metadata exist and are more likely to be updated.

Previous efforts similar to the ASCL demonstrated that coders are reluctant to maintain metadata on an external site \citep{RJH_adass1994,HEP_aas1998,CB_2004} and that software authors generally will not add their codes to such a service for many reasons, including not knowing the resource existed and not wanting to make an effort for something that may not have had much payback. We hope over time that more authors will add their codes, but realize there have to be good reasons for them to make the effort to do so; we hope to give them those good reasons.

We do not have DOIs at this time because of cost, but do have a unique identifier (the ASCL ID) for codes that are vetted by an editor and indexed by ADS. ADS can link to the software entry because of this identifier, and pick up citations using it if the citations are properly formatted.

\section{Licensing and Code Quality}
How a code is licensed doesn't matter to us. Codes do not need to be open source to be registered in the ASCL; any licensing is okay. We have GPL, Creative Commons, copyrighted software, open source, and everything in between. We do not care if programs are messy; if a code generated results, was used in research, we want it. And we do not care how software is cited, but do care that is  discoverable and cited. 

\section{Implementation and Operations}
The current infrastructure uses a mix of products. The code entries themselves are on a phpBB; our front end and blog use WordPress. We use Excel for building ADS upload files and a link checker written in Python. Finding codes is not as big a problem as finding all the necessary elements for inclusion in the ASCL, such as a research paper that uses a code or a code download site. We have dozens of code entries written up which need only one or the other of these elements to be moved into production. And we lack both time and money, as the effort is volunteer and unfunded.

\section{Maintaining a Balance}

The ASCL has to be lean enough to be managed and usable with volunteer effort to ensure sustainability over the long term. We need enough metadata to identify the code accurately and without ambiguity, and to link to it. Because the ADS indexes ASCL entries, information used by ADS remains static. Information that may change over time, such as version number, is not currently tracked in the ASCL; the curation effort to keep such metadata accurate is prohibitive. Complex searching that depends on more robust metadata cannot be done, either, since we don't have that metadata.

We weigh everything we do against the goal of code discoverability; we defer implementing other desirable features because of our resource constraints. 

Though we cannot do everything, we can keep the ASCL functional and usable by increasing the number of codes and maintaining the links to code sites. We run the link checker weekly; links that fail are investigated and if necessary replaced. We want to coordinate with journals to provide better exposure of the underlying computational methods of published research, and we want researchers to know about the ASCL.

In late 2011, we started a campaign that includes speaking at conferences, emailing code authors, and using links from APOD to the ASCL to increase awareness of the resource, and the ASCL has been advocating code release by participating in workshops and writing articles \citep{LS_AC2013}. We want code authors to receive recognition for their work, given how important software is to research and how much science it enables \citep{BW_2009astro}; the ASCL provides a way to cite a code without requiring a paper be written about the code. It also can serve as a repository for codes whose authors do not wish to maintain a website for their programs.

\section{The ASCL's Future}

Eventually, we would like to have a new infrastructure for code entries with additional features such as an API, better search capability, more professional look, and increased automation of the back end. 

The ASCL has some features we want to retain in a new infrastructure. It has a full-text iterative search, people can post comments by replying to a code entry thread, and the forum allows people to subscribe to a thread or the ASCL forum itself and be notified if there are additions to either. 

Despite its resource constraints, the ASCL continues to grow in the number of codes registered and its use by the community.

\bibliography{O29}










\end{document}